\documentclass[journal=jpclcd,manuscript=letter]{achemso}
\usepackage{amsmath,amssymb}
\usepackage{float}

\usepackage[T1]{fontenc}
\usepackage[utf8]{luainputenc}
\usepackage{float}
\usepackage{multirow}
\usepackage{lipsum}
\usepackage{bigints}
\usepackage{mathrsfs}
\usepackage{setspace}
\usepackage{droidsans}
\usepackage{balance}
\usepackage{times,mathptmx}
\usepackage{braket}
\usepackage{sectsty}
\usepackage{graphicx}
\usepackage{amsmath,amssymb}
\usepackage{mathrsfs}
\usepackage{multirow} 
\usepackage{lastpage}
\usepackage{adjustbox}
\usepackage{upgreek}
\usepackage{color}
\usepackage{array}
\usepackage{float}
\usepackage{fancyhdr}
\usepackage{fnpos}
\usepackage{adjustbox}
\usepackage[english]{babel}
\DeclareUnicodeCharacter{0308}{~}
\usepackage[version=3]{mhchem} 

\author{Preeti Bhumla}
\email{preeti.bhumla@physics.iitd.ac.in[PB]}
\author{Manjari Jain}
\author{Sajjan Sheoran}
\author{Saswata Bhattacharya}
\email{saswata@physics.iitd.ac.in[SB]}
\phone{+91-11-2659 1359}
\affiliation[Indian Institute of Technology Delhi]
{Department of Physics, Indian Institute of Technology Delhi, Hauz Khas, New Delhi 110016, India}
\title[An \textsf{achemso} demo]
{Vacancy-ordered Double Perovskites Cs$_2$BI$_6$ (B = Pt, Pd, Te, Sn): An Emerging Class of Thermoelectric Materials$^\dag$}
 \begin{document}

\begin{abstract}
 Vacancy-ordered double perovskites (A$_2$BX$_6$), being one of the environmentally friendly and stable alternatives to lead halide perovskites, have garnered considerable research attention in the scientific community. However, their thermal transport has not been explored much despite their potential applications. Here, we explore Cs$_2$BI$_6$ (B = Pt, Pd, Te, Sn) as potential thermoelectric materials using the state-of-the-art first-principles based methodologies, viz., density functional theory combined with many-body perturbation theory (G$_0$W$_0$) and spin-orbit coupling. 
The absence of polyhedral connectivity in vacancy-ordered perovskites gives rise to additional degrees of freedom leading to lattice anharmonicity. The presence of anharmonic lattice dynamics leads to strong electron-phonon coupling, which is well captured by Fr\"{o}hlich mesoscopic model. 
The lattice anharmonicity is further studied using {\it ab initio} molecular dynamics and electron localization function. The maximum anharmonicity is observed in Cs$_2$PtI$_6$, followed by Cs$_2$PdI$_6$, Cs$_2$TeI$_6$ and Cs$_2$SnI$_6$. Also, the computed average thermoelectric figure of merit ($zT$) for Cs$_2$PtI$_6$, Cs$_2$PdI$_6$, Cs$_2$TeI$_6$ and Cs$_2$SnI$_6$ are 0.88, 0.85, 0.95 and 0.78, respectively, which reveals their promising renewable energy applications.
  \begin{tocentry}
  \begin{figure}[H]%
  	\includegraphics[width=1.0\columnwidth,clip]{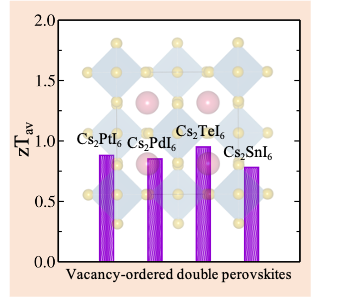}
  \end{figure}	
  \end{tocentry}
\end{abstract}
 Sustainable and renewable energy sources have become a long-standing aim in fulfilling the shortage of energy globally\cite{ajia2017generated, kojima2009organometal, burschka2013sequential, zhou2014interface}. Apart from solar energy, waste heat energy converted into electrical energy is a prominent source of renewable energy. To utilize this waste heat effectively, we need efficient thermoelectric materials~\cite{tan2016rationally, jing2019superior, mukhopadhyay2019lone}. Advantageously, the thermoelectric generators are durable, robust, scalable, compact and do not contain any moving parts. Note that, to have the maximum thermoelectric figure of merit ({\it zT}, see below), the material must have a large Seebeck coefficient ($S$) along with high electrical conductivity ($\sigma$) and a low thermal conductivity ($\kappa$)\cite{goldsmid2010introduction, zhu2017compromise, xie2020anomalously}. 
 \begin{equation}
 	zT = \frac{S^2 \sigma T}{\kappa}
 \end{equation}
 However, the strong coupling between these parameters with a trade-off relationship is challenging to achieve a high {\it zT} in a single system.
 
 Halide-based perovskites have unveiled a paradigm shift in the quest for high-performance materials\cite{bhumla2021origin, kojima2009organometal, ke2019unleaded, stoumpos2013semiconducting, arya2022rashba}. This can be attributed to their compositional and structural diversity that enables a wide array of functional properties~\cite{maughan2016defect}. More recently, halide perovskites have attracted attention for thermoelectric energy conversion due to their unique structural features and lattice dynamics\cite{acharyya2020intrinsically, lee2017ultralow, xie2020all, haque2020halide, jin2019hybrid}. Yang \textit{et al.}\cite{lee2017ultralow} reported an ultralow thermal conductivity of 0.5 Wm$^{-1}$K$^{-1}$ in halide perovskite nanowires composed of CsPbI$_3$, CsPbBr$_3$ and CsSnI$_3$. Additionally, several strategies viz., introduction of lattice defects in the structure\cite{kumar2021engineering, kumar2023effect}, creation of artificial superlattices and preparation of composite materials\cite{kumar2022synergistic}, have been explored to reduce the lattice thermal conductivities in materials. Most of the reported thermoelectric materials, such as SnSe\cite{zhao2014ultralow, zhao2016ultrahigh}, PbTe\cite{biswas2012high, pei2011self, pei2011convergence, yang2010high}, Cu$_2$Se\cite{bailey2016enhanced} and BiCuSeO\cite{li2015dual, yang2016manipulating}, have low intrinsic lattice thermal conductivity, similar in magnitude to that observed in halide perovskites. This exceptionally low thermal conductivity of halide perovskites, in conjunction with their high carrier mobility, makes them promising for thermoelectric investigations\cite{xie2020all, qian2016lattice, saxena2016enhanced}. However, unfortunately, these alluring materials suffer from lead toxicity and long-term instability. These drawbacks have motivated the scientific community to explore alternative perovskite compositions and structures.
 
\begin{figure}[h]
	\includegraphics[width=0.98\textwidth]{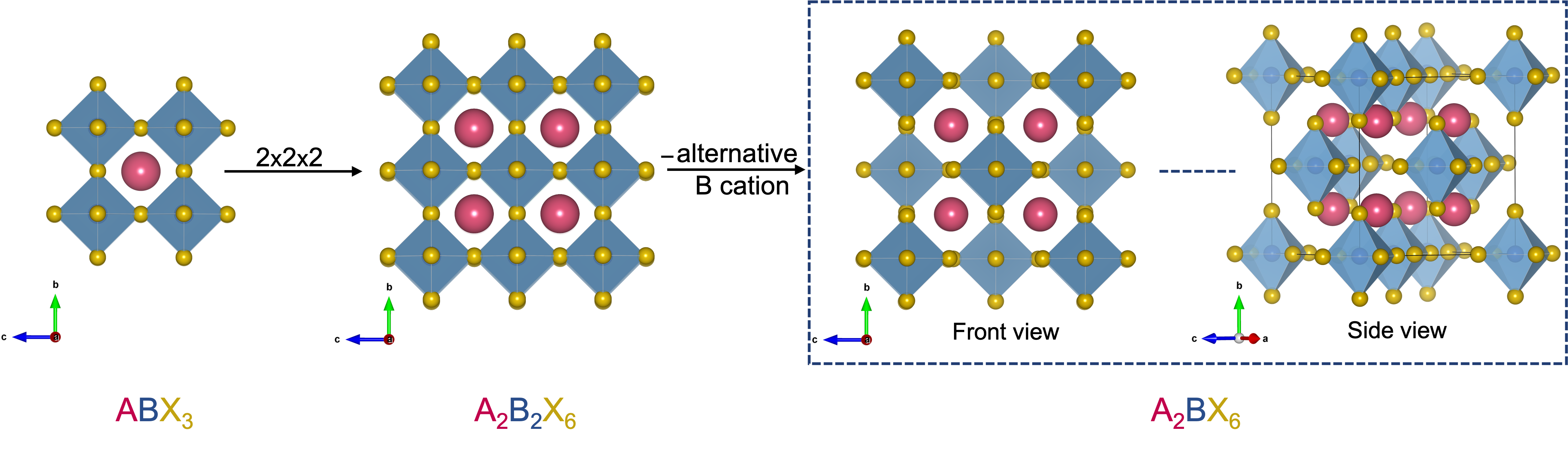}
	\caption{Schematic of the relationship between ABX$_3$ perovskite and A$_2$BX$_6$ (vacancy-ordered) double perovskite.}
	\label{pic1}
\end{figure}

 One alternative class of materials is the inorganic lead-free double perovskites with the general formula A$_2$BX$_6$, commonly known as vacancy-ordered double perovskites. This defect-variant of halide perovskite is derived by doubling the ABX$_3$ unit cell along all three crystallographic axes and removing every alternate B-site cation, as illustrated in Figure \ref{pic1}. These perovskites provide new opportunities for non-toxic and stable replacements of Pb and Sn. Lately, vacancy-ordered double perovskites have been explored in thermoelectrics due to their ultralow lattice thermal conductivity, which is attributed to their highly anharmonic lattice dynamics~\cite{acharyya2020intrinsically, xie2020all, klarbring2020anharmonicity}. More recently, Jong \textit{et al.} have reported that twofold rattling modes of Cs atoms and SnI$_6$ clusters lead to ultralow thermal conductivity in Cs$_2$SnI$_6$\cite{jong2022twofold}.

Motivated by this idea,  in this Letter, we have studied the vacancy-ordered double perovskites Cs$_2$BI$_6$ (B = Pt, Pd, Te, Sn) using state-of-the-art first-principles based methodologies under the framework of density functional theory (DFT)\cite{hohenberg1964inhomogeneous, kohn1965self} with suitable exchange-correlation ($\epsilon_{xc}$) functionals combined with many-body perturbative approaches (G$_0$W$_0$)~\cite{hedin1965new, hybertsen1985first} and spin-orbit coupling (SOC). First, we have examined the structural, mechanical and thermodynamic stability of these materials. After that, we have studied the thermoelectric properties, where we find reasonably high $zT$ values, calculated as a function of temperature ($T$). Interestingly, we have observed the presence of anharmonic effects, which are quantified by computing the harmonic and anharmonic energy contribution as a function of temperature in these vacancy-ordered double perovskites. In order to further explore the underlying mechanisms, we have examined the dynamical stability and anharmonicity by computing the phonon bandstructures and electron localization function (ELF). The ELF further confirms the presence of anharmonicity in this class of systems. As a consequence, we find that the study of electron-phonon interaction is important. The electron-phonon interaction is well captured by studying Fr\"{o}hlich mesoscopic model\cite{frost2017calculating, frohlich1954electrons} to investigate the interaction of longitudinal optical phonon modes with the carriers that strongly influence the carrier mobility. 

Cs$_2$BI$_6$ (B = Pt, Pd, Te, Sn) vacancy-ordered double perovskites have a face-centered cubic crystal structure with the space group $Fm$$\overline{3}$$m$ (no. 225). In this structure, the Cs atoms are situated at the 8c Wyckoff positions and (0.25, 0.25, 0.25) coordinates, B atoms at 4a Wyckoff positions and (0, 0, 0) coordinates, and I atoms at 24e Wyckoff positions and ($x$, 0, 0) coordinates, where the value of $x$ is around 0.20. Each of the Cs atoms resides between the [BI$_6$] octahedra and is surrounded by 12 I atoms, while the B atoms are at the corners and face-centered positions of [BI$_6$] octahedra. The Cs atoms located in the octahedral cage can act as heavy rattlers and lead to lattice anharmonicity\cite{jong2022twofold}. The optimized lattice parameters of Cs$_2$BI$_6$ perovskites are listed in Table S1 of Supporting Information (SI). 

Firstly, we have examined the stability of material as it is an essential factor in achieving a high-performance device applications. In order to predict the structural stability of the vacancy-ordered double perovskites, we have calculated the Goldschmidt tolerance factor ($t$)\cite{goldschmidt1926gesetze, sun2017thermodynamic} and octahedral factor ($\mu$)\cite{travis2016application}. Recently, a new tolerance factor ($\tau$)\cite{bartel2019new}, proposed by Bartel {\it et al.} is also computed and compared for all the considered perovskites (see details in Section II of SI). The calculated values show that these perovskites are stable in cubic structures. Besides structural stability, we have also calculated the thermodynamic and mechanical stabilities. For thermodynamic stability\cite{PhysRevB.99.184105, jong2018first}, we have calculated the Gibbs free energies ($\Delta G$) of perovskites as per the equations given in Section III of SI. 
 $\Delta G$ is calculated using both Perdew-Burke-Ernzerhof (PBE)\cite{PhysRevLett.77.3865} and Heyd–Scuseria–Ernzerhof (HSE06)~\cite{heyd2003hybrid, krukau2006influence} $\epsilon_{xc}$ functionals in the Cs$_2$BI$_6$ perovskites. The values of $\Delta G$ are negative, indicating the thermodynamic stability of these perovskites (for details, refer Table S3 of SI). 

Subsequently, to determine the mechanical stability of the perovskites, we have calculated the elastic constants of the materials using the finite strain theory\cite{murnaghan1937finite}. For cubic symmetry, three independent elastic constants viz., $C_{11}$, $C_{12}$ and $C_{44}$ are sufficient to explain the mechanical stability and related properties of the crystal. Using these elastic constants, we can calculate Bulk ($B$), Shear ($G$) and Young’s modulus ($E$) of the perovskites (for details, see Section IV of SI). The corresponding mechanical stability criterion\cite{PhysRevB.63.134112, mouhat2014necessary} is given as follows:
\begin{equation}
	C_{11} > 0,  \: \: C_{44} > 0, \: \: C_{11} - C_{12} > 0, \: \:  C_{11} + 2C_{12} > 0
\end{equation}



The calculated elastic constants and moduli are given in Table 1. As we can see, the elastic constants satisfy the stability criteria, indicating the mechanical stability of these vacancy-ordered double perovskites. The fragility of these perovskites are studied in terms of Pugh's ($B/G$) and Poisson's ratio ($\nu$). 
The calculated values of $B/G$ (> 1.75) and $\nu$ (> 0.26) show that the studied vacancy-ordered double perovskites are ductile (see Table 1 below). 
Also, we have calculated the elastic anisotropy ($A$) of these materials, given by the equation:
\begin{equation}
	A=\frac{2C_{44}}{C_{11}-C_{12}}
\end{equation}
where $A$ represents the elastic anisotropy coefficient. 
The value of $A$ is equal to 1 for an isotropic crystal. The deviation from this value measures the degree of elastic anisotropy possessed by the crystal. According to the calculated values, all the considered double perovskites are anisotropic in nature.
\begin{table}[htbp]
	\caption {Calculated elastic constants C$_{ij}$ (GPa), Bulk modulus $\mathbf {\textit B}$ (GPa), Shear modulus $\mathbf {\textit G}$ (GPa), Young's modulus $\mathbf { \textit E}$ (GPa), Pugh's ratio $\mathbf {\textit{B/G}}$, Poisson's ratio \textbf{$\nu$} and elastic anisotropy \textbf{\textit{A}} of Cs$_2$BI$_6$ vacancy-ordered double perovskites.}
	\begin{center}
		\begin{adjustbox}{width=0.75\textwidth} 
			\setlength\extrarowheight{+4pt}
			\begin{tabular}[c]{|c|c|c|c|c|c|c|c|c|c|} \hline		
				\textbf{Configurations} & \textbf{$\textit{C}_{11}$} & \textbf{$\textit{C}_{12}$} & \textbf{$\textit{C}_{44}$} & \textbf{\textit{B}} & \textbf{\textit{G}} & \textbf{\textit{E}} & \textbf{\textit{B/G}} & \textbf{$\nu$} &  \textbf{\textit{A}} \\ \hline
				Cs$_2$PtI$_6$        & 9.58  & 4.51 & 3.93 &  6.20    &     3.30 & 8.40  & 1.88 & 0.27 &  1.55 \\  \hline
				Cs$_2$PdI$_6$        &  16.64  &  8.98  & 7.36 & 11.53    &     5.66    & 11.39 & 2.04  & 0.29 & 1.92 \\ \hline
				Cs$_2$TeI$_6$        & 20.30  & 10.55 & 8.70 & 13.80     &     6.90  &  17.74  & 2.00 & 0.29 & 1.78\\ \hline
				Cs$_2$SnI$_6$       &  14.36  &  8.20  & 6.65 & 10.25     &      4.88    &  12.63 & 2.10  & 0.29 & 2.16 \\ \hline
			\end{tabular}
		\end{adjustbox}
		\label{T1}
	\end{center}
\end{table}

After studying the stability, we have calculated the electronic band gaps (with and without SOC) of the vacancy-ordered double perovskites. Since simple local/semi-local $\epsilon_{xc}$ functionals (viz., LDA, GGA) are unable to predict the band gaps correctly due to their incapability of capturing the electron's self-interaction error, we have employed HSE06 $\epsilon_{xc}$ functional and many-body perturbation theory (G$_0$W$_0$) to calculate the band gaps more accurately. The calculated band gaps of Cs$_2$PtI$_6$, Cs$_2$PdI$_6$, Cs$_2$TeI$_6$ and Cs$_2$SnI$_6$ are 1.35, 1.43, 1.49 and 1.23 eV, respectively. This implies that all these perovskites have  band gaps in the visible region, expanding their scope for energy-harvesting applications. The values of band gaps are listed in Table \ref{T2}, which agree well with the experimental values. The bandstructures, projected density of states (pDOS) and $k$-grid convergence data are provided for all perovskites in Section V-VII of SI.

\begin{table}[htbp]
	\caption {Band gap (eV) of Cs$_2$BI$_6$ vacancy-ordered double perovskites calculated using different $\epsilon_{xc}$ functionals.}
	\begin{center}
		\begin{adjustbox}{width=1.0\textwidth} 
			\setlength \extrarowheight {+5pt}
			\begin{tabular}[c]{|c|c|c|c|c|c|c|c|} \hline		
				\textbf{Configurations} & \textbf{PBE} & \textbf{PBE+SOC} & \textbf{HSE06} & \textbf{HSE06+SOC} & \textbf{G$_0$W$_0$@PBE+SOC} &  \textbf{G$_0$W$_0$@HSE06+SOC} & \textbf{Experimental}\\ \hline
				Cs$_2$PtI$_6$        &  0.36  & 0.29 & 1.07 & 0.96  & \textbf{1.35}  & 2.20  & 1.37\cite{yang2020novel} \\ \hline
				Cs$_2$PdI$_6$       & 0.06   &  0.02  &  0.62  &    0.51    &  0.59 & \textbf{1.43}& 1.41\cite{zhou2018all} \\ \hline
				Cs$_2$TeI$_6$       &  1.14  & 0.91 & 1.70 & \textbf{1.49}    &    2.12 & 2.44 & 1.50\cite{vazquez2020vacancy} \\ \hline
				Cs$_2$SnI$_6$       &  0.09  &  0.06  & 0.84 & 0.70  &  \textbf{1.23} & 2.31 &1.25\cite{maughan2016defect}\\ \hline
			\end{tabular}
		\end{adjustbox}
		\label{T2}
	\end{center}
\end{table}

\begin{figure}[h]
	\includegraphics[width=0.98\textwidth]{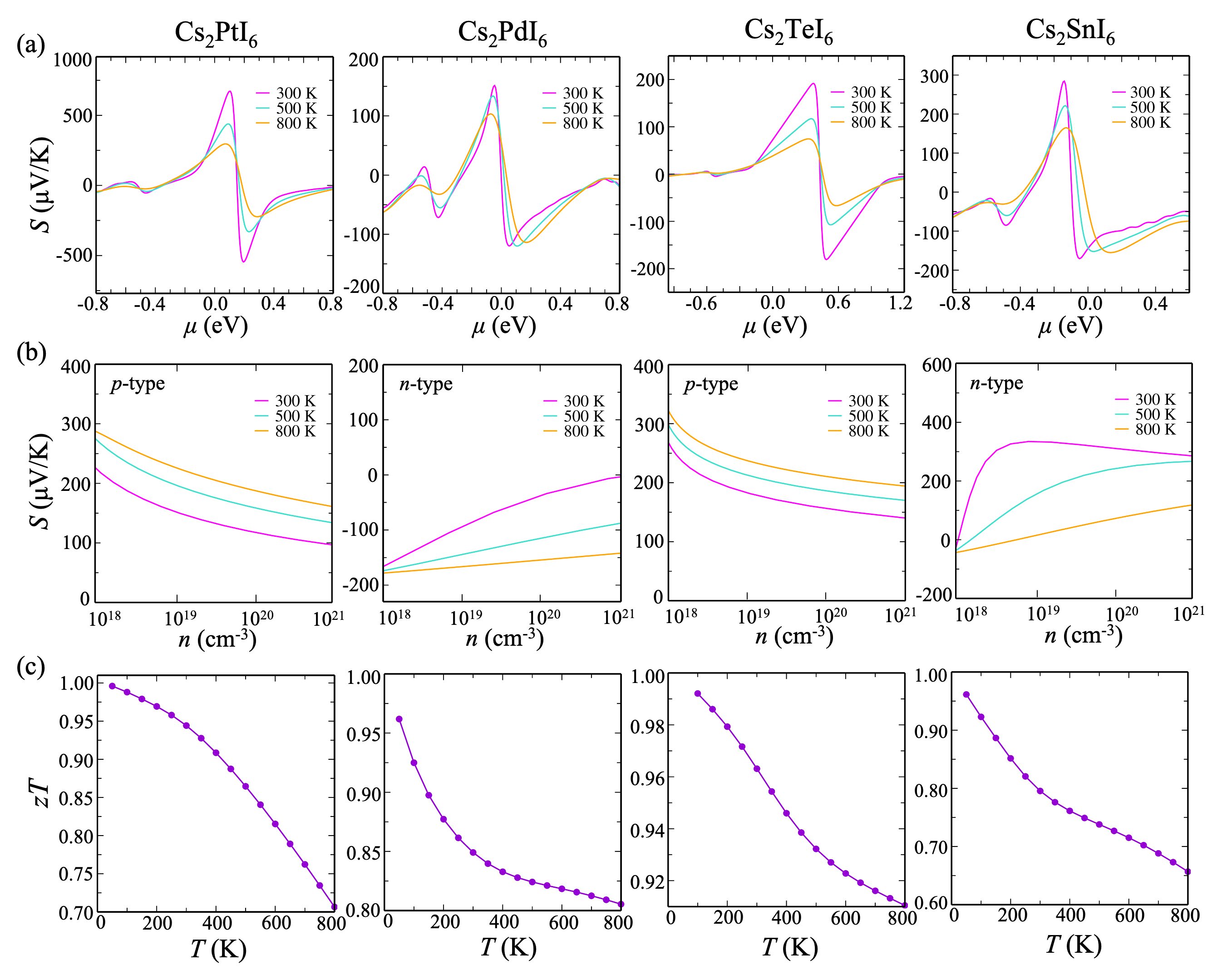}
	\caption{Seebeck coefficient ($S$) as a function of (a) chemical potential ($\mu$) and (b) carrier concentration ($n$) at 300, 500 and 800\:K. (c) Thermoelectric figure of merit ($zT$) as a function of temperature ($T$) for Cs$_2$PtI$_6$, Cs$_2$PdI$_6$, Cs$_2$TeI$_6$ and Cs$_2$SnI$_6$, calculated using HSE06 $\epsilon_{xc}$ functional.}
	\label{pic10}
\end{figure}

Next, we have calculated the thermoelectric properties of the perovskites using the BoltzTrap Code\cite{madsen2006boltztrap}. Figure \ref{pic10}(a) shows the Seebeck coefficient ($S$) as a function of chemical potential ($\mu$) for Cs$_2$BI$_6$ perovskites (for computational details, see Section VIII of SI). $S$ measures the induced thermoelectric voltage ($\Delta$$V$) in response to a temperature difference ($\Delta$$T$) across the material and is given as $S$=$\Delta$$V$/$\Delta$$T$, while $\mu$ shows the addition or removal of electrons (doping) against the repulsive forces of electrons already present in the material. Advantageously, the position of $\mu$ determines the fraction of electrons in the conduction or valence band which take part in the electronic transport and hence influences $S$. Thus, $S$ can be determined from the change in the chemical potential of electrons induced by the temperature difference. By definition, $\mu$=0 coincides with the top of the valence band in semiconductors. This implies that at $\mu$=0, the nature of $S$ determines the type of semiconductor. From Figure \ref{pic10}(a), we can see that at $\mu$=0, the value of $S$ is positive for Cs$_2$PtI$_6$ and Cs$_2$TeI$_6$ at various temperatures, indicating that these perovskites are $p$-type semiconductors. The maximum values of $S$ for Cs$_2$PtI$_6$ and Cs$_2$TeI$_6$ are 710 and 190 \textmu V/K, respectively, at 300\:K. On the other hand, $S$ is negative for Cs$_2$PdI$_6$ and Cs$_2$SnI$_6$, which indicates the $n$-type character in these perovskites\cite{bhui2022intrinsically}. For Cs$_2$PdI$_6$ and Cs$_2$SnI$_6$, the maximum values of $S$ are 148 and 290 \textmu V/K, respectively, at 300\:K. Also, we have observed that the value of $S$ decreases with an increase in temperature for all the considered Cs$_2$BI$_6$ perovskites. Figure \ref{pic10}(b) shows the variation of $S$ with carrier concentration ($n$)\cite{sajjad2020ultralow, faizan2021electroni} at different temperatures. Subsequently, to calculate the efficiency of material to convert heat into electrical energy, we have calculated the $zT$ as a function of temperature (see Figure \ref{pic10}(c)) for Cs$_2$BI$_6$ perovskites. The computed average $zT$ for Cs$_2$PtI$_6$, Cs$_2$PdI$_6$, Cs$_2$TeI$_6$ and Cs$_2$SnI$_6$ are 0.88, 0.85, 0.95 and 0.78, respectively, which make them promising for thermoelectric applications.

\begin{figure}[h]
	\includegraphics[width=0.99\textwidth]{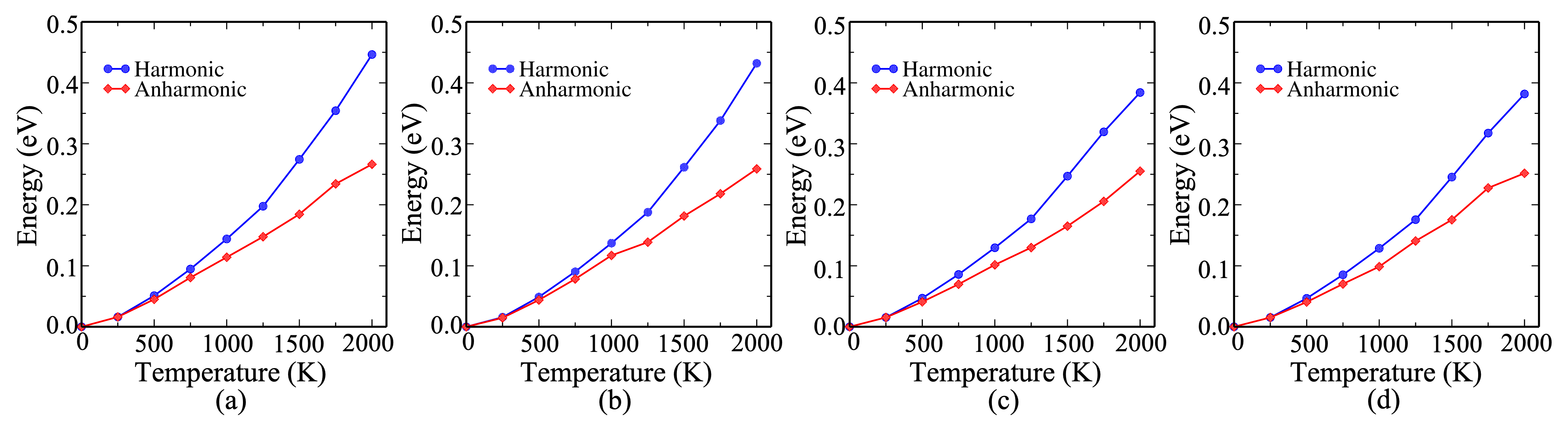}
	\caption{Harmonic (U$_{h}$) and anharmonic energy (U$_{ah}$) of (a) Cs$_2$PtI$_6$ (b) Cs$_2$PdI$_6$ (c) Cs$_2$TeI$_6$ and (d) Cs$_2$SnI$_6$ as a function of temperature ($T$), calculated using HSE06 $\epsilon_{xc}$ functional.}
	\label{pic4}
\end{figure}

Low thermal conductivity is desirable for efficient thermoelectric materials, which in turn depends on lattice dynamics\cite{graff2014reduced, chen2011origin}. Lattice dynamics play a pivotal role in governing materials properties such as thermal conductivity\cite{dugdale1955lattice}, ionic and electronic transport\cite{muy2018tuning}, optical emission\cite{chodos1976vibronic}, ferroelectricity and superconductivity\cite{bardeen1957theory}. Deviation from harmonic vibrational potential results in high amplitude anharmonic vibrations that introduce vibrational disorder in the system. This results in significant phonon-phonon scattering, which leads to low thermal conductivities and better thermoelectric performance\cite{jana2018crystalline, katsnelson2005lattice, paul2021electron} (see details in Section IX of SI). 
To examine this deviation at high temperatures, we have calculated the harmonic (U$_{h}$) as well as anharmonic energy (U$_{ah}$) in vacancy-ordered double perovskites. U$_{h}$ is calculated under harmonic approximation for all perovskites. To quantify U$_{ah}$ in Cs$_2$BI$_6$ vacancy-ordered double perovskites, we have performed AIMD calculations at different temperatures using Nose-Hoover thermostat~\cite{evans1985nose}. This data is then fed to a post-processing python package pyHMA\cite{moustafa2021pyhma}, which determines the anharmonic energy (for details, see Section X of SI). Figure \ref{pic4}(a-d) show the variation of U$_{ah}$ with temperature. As the temperature increases, we observe a deviation from harmonic potential leading to lattice anharmonicity. 

\begin{figure}[h]
	\includegraphics[width=0.99\textwidth]{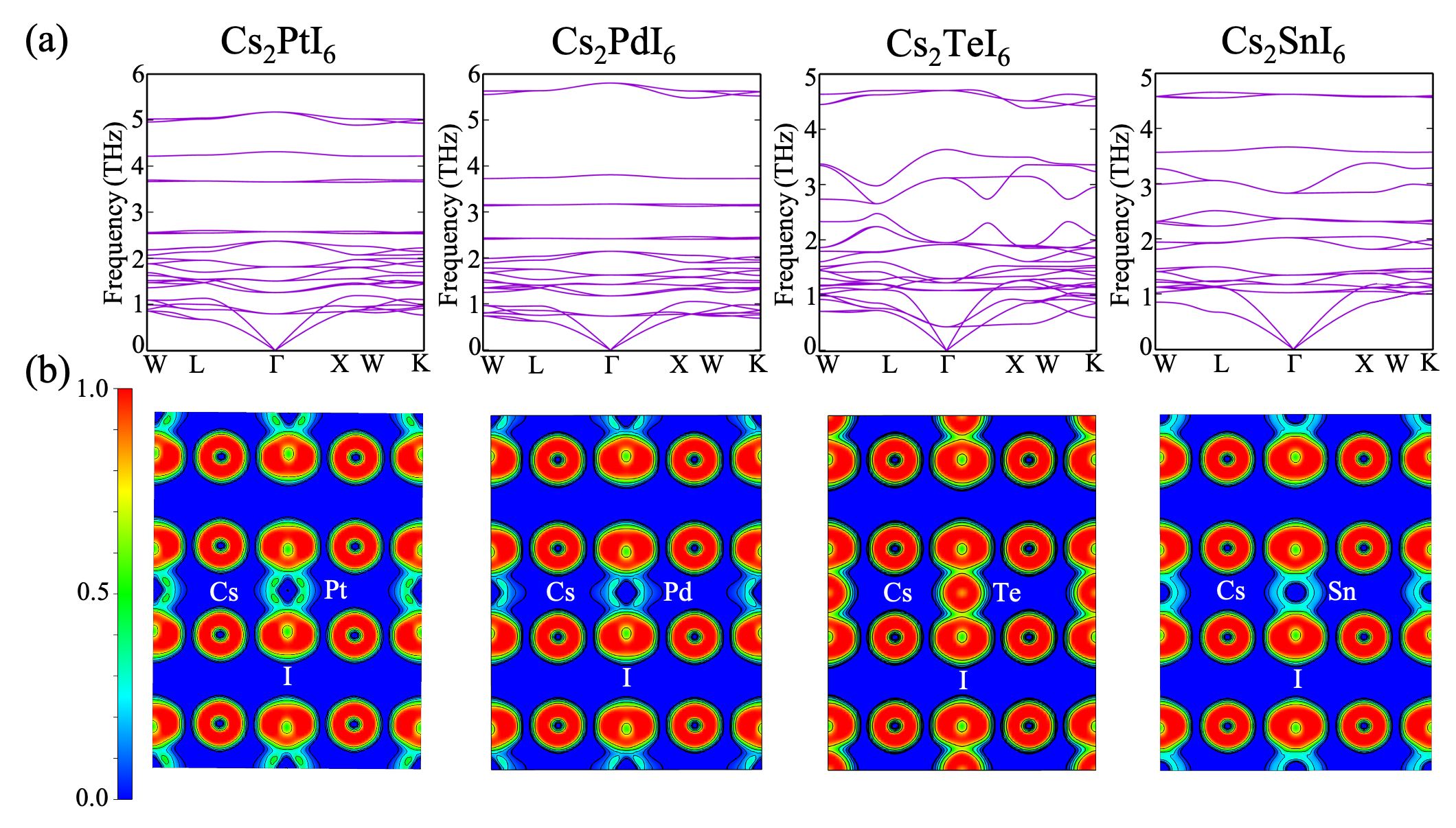}
	\caption{(a) Phonon dispersion plots and (b) two dimensional ELF of (110) plane passing through Cs, Pt, Pd, Te, Sn and I atoms of Cs$_2$PtI$_6$, Cs$_2$PdI$_6$, Cs$_2$TeI$_6$ and Cs$_2$SnI$_6$, calculated using PBE $\epsilon_{xc}$ functional. The ELF values of 0.0, 0.5, and 1.0 are interpreted as absence, uniform electron gas like, and localized electrons, respectively.}
	\label{pic2}
\end{figure}

The intimate connection between anharmonic lattice dynamics and functional properties motivates a fundamental understanding of anharmonicity in this class of materials. To further assess anharmonicity in our system, we have first examined the dynamical stability by plotting the phonon dispersion bandstructures of all Cs$_2$BI$_6$ perovskites using density functional perturbation theory (DFPT)\cite{gajdovs2006linear}, as shown in Figure \ref{pic2}(a). For vacancy-ordered double perovskites, the structural symmetry confirms 108 phonon modes as they contain 36 atoms per unit cell. Out of these 108 phonon modes, there are 3 acoustic modes, while the remaining modes are optical, characterized as low and high-frequency phonons, respectively. The absence of negative frequencies confirms the dynamical stability of these perovskites. After examining the phonon modes, we try to explore the interaction between the atoms of these perovskites. The spatial distribution of the electron density around atom gives the measure of phonon anharmonicity. Therefore, we have computed the ELF to study materials bonding and anharmonicity (see Figure \ref{pic2}(b)). The localization of electrons is estimated by a dimensionless ELF probability density ranging between 0 and 1. With the increase in ELF value, the electrons get more localized, and hence the bonds become stronger. As we can see in Figure \ref{pic2}(b), I atoms draw more charge because of its higher electronegativity in comparison to B atoms. Nevertheless, there is significant charge sharing among B-I bonds due to the small electronegativity difference, indicating the possibility of covalent bonding. On the other hand, no charge is shared between Cs and B/I atoms. However, physical interaction between Cs and [BI$_6$] octahedra results in nonspherical electron density around Cs and I atoms, which explains the origin of the phonon anharmonicity.  Also, there is no charge transfer between the neighbouring octahedra owing to the vacancies in Cs$_2$BI$_6$ perovskites. This indicates that [BI$_6$] octahedra in Cs$_2$BI$_6$ are loosely bound and may lead to cluster-rattling vibrations along with Cs atom rattlers\cite{jong2022twofold}. This in turn, increases the phonon scattering followed by suppression of $\kappa$\cite{katsnelson2005lattice, paul2021electron}. 

Anharmonic lattice dynamics give rise to stronger electron-phonon coupling in the material as electrons interact with lattice vibrations via the formation of polarons\cite{wang2017dynamical, leveillee2021phonon}. To study these electron-phonon interactions, we have calculated the electron-phonon coupling strength using Fr\"{o}hlich’s polaron model\cite{frohlich1954electrons}. The dimensionless Fr\"{o}hlich electron-phonon coupling parameter ($\alpha$) measures the electron-phonon coupling strength of the material and is given as:
\begin{equation}
	\begin{split}
		\alpha = \frac{1}{4\pi\epsilon_0}\frac{1}{2}\left(\frac{1}{\epsilon_\infty}-\frac{1}{\epsilon_\textrm{static}}\right)\frac{\textrm{e}^2}{\hbar\omega_{\textrm{LO}}}\left({\frac{2{m}^*\omega_{\textrm{LO}}}{\hbar}}\right)^{1/2}
		\label{eq1}
	\end{split}
\end{equation}
The $\alpha$ depends on the material-specific properties, viz., electronic ($\epsilon_{\infty}$) and ionic static ($\epsilon_{\textrm{static}}$) dielectric constants, permittivity of free space ($\epsilon_0$), the effective carrier mass (${m}^*$) (for calculation of effective mass, see Section XI in SI), and a characteristic longitudinal optical phonon angular frequency ($\omega_{\textrm{LO}}$). For a system having multiple phonon branches, an average LO frequency is calculated by considering all the infrared active optical phonon branches and taking a spectral average of them\cite{hellwarth1999mobility} (for details of calculation of $\omega_{\textrm{LO}}$ and different polaron parameters, see Section XII of SI).
\begin{table}[h]
	\caption {Effective mass of electrons \textit{m}$^{*}$ in terms of rest mass of electron \textit{m}$_0$, electronic dielectric constant $\epsilon_\textrm{$\infty$}$ and ionic static dielectric constant $\epsilon_\textrm{static}$ of Cs$_2$BI$_6$ perovskites.}
	\begin{center}
		\begin{adjustbox}{width=0.4\textwidth} 
			\setlength\extrarowheight{+4pt}
			\begin{tabular}[c]{|c|c|c|c|} \hline		
				\textbf{Configurations} & \textbf{\textit{m}$^{*}$} & \textbf{$\epsilon_\textrm{$\infty$}$} & \textbf{$\epsilon_\textrm{static}$} 
				\\ \hline
				Cs$_2$PtI$_6$    &  0.49    &   3.41 &  4.55  \\  \hline
				Cs$_2$PdI$_6$   &  0.47    &  2.80  & 3.30   \\ \hline
				Cs$_2$TeI$_6$   &  0.40    &  3.59 &  5.55 \\ \hline
				Cs$_2$SnI$_6$   &  0.33   &   2.67  & 3.46   \\ \hline
			\end{tabular}
		\end{adjustbox}
		\label{T1}
	\end{center}
\end{table}
The values of $\alpha$ for Cs$_2$BI$_6$ perovskites are listed in Table \ref{T1} and follow the order: Cs$_2$PtI$_6$ > Cs$_2$PdI$_6$ > Cs$_2$TeI$_6$ > Cs$_2$SnI$_6$. The ELF and strong electron-phonon coupling validate the presence of anharmonicity in our considered perovskites. Also, the heavy atoms present in the system act as phonon rattlers and help suppress the lattice thermal conductivity effectively. This leads to their effective utilization in high-performance thermoelectric device applications.
\begin{table}[htbp]
	\caption {Calculated polaron parameters corresponding to electrons in Cs$_2$BI$_6$ perovskites. $\omega$$_{\textrm{LO}}$\:(THz), $\alpha$$_{\textrm{e}}$, \textit{m}$_{\textrm{P}}$, \textit{l}$_\textrm{P}$\:(\AA) and \textit{$\mu$}$_\textrm{P}$\:(cm$^2$V$^{-1}$s$^{-1}$) are the optical phonon frequency, Fr\"{o}hlich coupling constant, polaron mass, polaron radii and polaron mobility, respectively.}
	\begin{center}
		\begin{adjustbox}{width=0.80\textwidth} 
			\setlength\extrarowheight{+4pt}
			\begin{tabular}[c]{|c|c|c|c|c|c|} \hline		
				\textbf{Configurations} & \textbf{$\omega$$_{\textrm{LO}}$\:(THz)} &  \textbf{$\alpha_{\textrm{e}}$} & \textbf{{\textit{m}}$_\textrm{P}$/{\textit{m}}$^{*}$} & \textbf{\textit{l}$_\textrm{P}$\:(\AA)} & \textbf{{$\mu_\textrm{P}$}\:{\textrm{(cm$^2$V$^{-1}$s$^{-1}$)}}} \\ \hline
				Cs$_2$PtI$_6$   &  1.74  & 2.24 & 1.29  & 22.90  &   42.15 \\ \hline
				Cs$_2$PdI$_6$  &  1.62  & 1.81  & 1.19  &  25.10  &   64.90 \\ \hline
				Cs$_2$TeI$_6$  &  4.26 & 1.73 & 1.35  &  50.98 &  45.30 \\ \hline
				Cs$_2$SnI$_6$  &  3.52 & 1.50 & 1.27  &  47.43 &    71.12   \\ \hline
			\end{tabular}
		\end{adjustbox}
		\label{T1}
	\end{center}
\end{table}
\newpage
In summary, we have carried out an exhaustive study to investigate the structural, elastic and thermoelectric properties of Cs$_2$BI$_6$ (B = Pt, Pd, Te, Sn) vacancy-ordered perovskites under the framework of density functional theory. The Pugh's and Poisson's ratios show the ductile nature of the perovskites. Also, the negative Gibbs free energies and phonon bandstructures confirm the stability of these perovskites. The band gaps calculated using different $\epsilon_{xc}$ functionals appear in visible region, which is advantageous for energy-harvesting properties. 
The $zT$ values for Cs$_2$PtI$_6$, Cs$_2$PdI$_6$, Cs$_2$TeI$_6$ and Cs$_2$SnI$_6$ are 0.88, 0.85, 0.95 and 0.78, respectively, which show that these perovskites are promising for thermoelectric applications. To examine the role of anharmonicity, the ELFs are plotted for these perovskites, which indicates the presence of lattice anharmonicity. The calculation of the harmonic and anharmonic energy confirm the anharmonicity in this class of materials. As a result of anharmonicity, these compounds have strong electron-phonon coupling and the strength of this coupling is quantified using Fr\"{o}hlich’s polaron model. The phonon-phonon scattering owing to the presence of anharmonicity and heavy atoms acting as phonon rattlers results in low thermal conductivity and better thermoelectric properties.

\section{Computational Methods}
All the DFT\cite{hohenberg1964inhomogeneous, kohn1965self} calculations have been performed using the Vienna $ab$ $initio$ simulation package (VASP)\cite{kresse1993ab, kresse1996efficient}. The ion-electron interactions in all the elemental constituents are described using the projector augmented wave (PAW)\cite{blochl1994projector, kresse1999ultrasoft} method as implemented in VASP. The structural optimization is performed using generalized gradient approximation (PBE)\cite{PhysRevLett.77.3865} and optB86\cite{thonhauser2007van} $\epsilon_{xc}$ functional with vdW corrections, relaxing all ions until Hellmann-Feynman forces are less than 0.001 eV/\AA. The two-body vdW interaction, devised by Tkatchenko-Scheffler has been used during optimization\cite{tkatchenko2009accurate}. The cutoff energy of 600 eV is used for the plane wave basis set such that the total energy calculations are converged within 10$^{-5}$ eV. The $\Gamma$-centered 4$\times$4$\times$4 $k$-grid is used to sample the Brillouin zone except when stated otherwise. The band gap is calculated using hybrid $\epsilon_{xc}$ functional (HSE06)~\cite{heyd2003hybrid, krukau2006influence} and many-body perturbation theory. Note that the single-shot GW (G$_0$W$_0$)~\cite{hedin1965new, hybertsen1985first} calculations have been performed on top of the orbitals obtained from PBE/HSE06 (with SOC) $\epsilon_{xc}$ functional [G$_0$W$_0$@PBE+SOC/HSE06+SOC]. The number of bands is set to four times the number of occupied bands. The polarizability calculation is performed on a grid of 50 frequency points. The effective mass is calculated by SUMO\cite{ganose2018sumo} using a parabolic fitting of the band edges. The phonon calculation is performed for 2$\times$2$\times$2 supercell using the PHONOPY package\cite{togo2008first, togo2015first}. The BoltzTrap Code\cite{madsen2006boltztrap} based on Boltzmann transport theory and phono3py\cite{phono3py} are used to evaluate thermoelectric properties. The self-consistent process described by Hellwarth is used to calculate the electron-phonon coupling strength\cite{frost2017calculating}. Static dielectric constant is calculated using density functional perturbation theory (DFPT)\cite{gajdovs2006linear} with a denser $k$-grid (6$\times$6$\times$6). To calculate the anharmonic energy, we have carried out {\it ab initio} molecular dynamics (AIMD) simulation employing Nose-Hoover thermostat~\cite{evans1985nose} and pyHMA package\cite{moustafa2021pyhma}.
\begin{acknowledgement}
P.B. acknowledges UGC, India, for the senior research fellowship [grant no. 1392/(CSIR-UGC NET JUNE 2018)]. M.J. acknowledges CSIR, India, for the senior research fellowship [Grant No. 09/086(1344)/2018-EMR-I]. S.S. acknowledges CSIR, India, for the senior research fellowship [grant no. 09/086(1432)/2019-EMR-I].  S.B. acknowledges financial support from SERB under a core research grant (Grant No. CRG/2019/000647) to set up his High Performance Computing (HPC) facility “Veena” at IIT Delhi for computational resources.
\end{acknowledgement}
\begin{suppinfo}
See supplementary material for the details of optimized lattice parameters, Goldschmidt tolerance factor ($t$), octahedral factor ($\mu$) and new tolerance factor ($\tau$), Gibbs free energy ($\Delta G$), mechanical properties, bandstructures, projected density of states (pDOS), $k$-grid convergence, computational details of Seebeck coefficient ($S$) and thermoelectric figure of merit ($zT$), thermal conductivity ($\kappa$), harmonic (U$_{h}$) and anharmonic energy (U$_{ah}$), effective mass and polaron parameters of Cs$_2$BI$_6$ perovskites vacancy-ordered perovskites.
\end{suppinfo}

\providecommand{\latin}[1]{#1}
\makeatletter
\providecommand{\doi}
{\begingroup\let\do\@makeother\dospecials
	\catcode`\{=1 \catcode`\}=2 \doi@aux}
\providecommand{\doi@aux}[1]{\endgroup\texttt{#1}}
\makeatother
\providecommand*\mcitethebibliography{\thebibliography}
\csname @ifundefined\endcsname{endmcitethebibliography}
{\let\endmcitethebibliography\endthebibliography}{}

\end{document}


\begin{center}
{\Large \bf Supplemental Material}\\ 
\end{center}
\begin{enumerate}[\bf I.]
	
\item Lattice parameters of Cs$_2$BI$_6$ (B = Pt, Pd, Te, Sn) vacancy-ordered double perovskites
\item Goldschmidt tolerance factor ($t$), octahedral factor ($\mu$) and new tolerance factor ($\tau$) of Cs$_2$BI$_6$ (B = Pt, Pd, Te, Sn) vacancy-ordered double perovskites
\item Gibbs free energy ($\Delta G$) of Cs$_2$BI$_6$ (B = Pt, Pd, Te, Sn) vacancy-ordered double perovskites
\item Mechanical properties of Cs$_2$BI$_6$ (B = Pt, Pd, Te, Sn) vacancy-ordered double perovskites
\item Band structures of Cs$_2$BI$_6$ (B = Pt, Pd, Te, Sn) vacancy-ordered double perovskites
\item Projected density of states (pDOS) of Cs$_2$BI$_6$ (B = Pt, Pd, Te, Sn) vacancy-ordered double perovskites
\item $k$-grid convergence in Cs$_2$BI$_6$ (B = Pt, Pd, Te, Sn) vacancy-ordered double perovskites
\item Computational details for calculation of Seebeck coefficient ($S$) and thermoelectric figure of merit ($zT$) in Cs$_2$BI$_6$ (B = Pt, Pd, Te, Sn) vacancy-ordered double perovskites
\item Thermal conductivity ($\kappa$) of Cs$_2$BI$_6$ (B = Pt, Pd, Te, Sn) vacancy-ordered double perovskites
\item Calculation of harmonic (U$_{h}$) and anharmonic energy (U$_{ah}$) in Cs$_2$BI$_6$ (B = Pt, Pd, Te, Sn) vacancy-ordered double perovskites
\item Effective mass calculation of Cs$_2$PtI$_6$ vacancy-ordered double perovskite
\item Calculation of longitudinal optical phonon frequency ($\omega_{\textrm{LO}}$), effective mass of polaron (\textit{m}$_\textrm{P}$), polaron radii (\textit{l}$_\textrm{P}$) and polaron mobility ($\mu_\textrm{P}$)
\end{enumerate}
\vspace*{12pt}
\clearpage

\section{Lattice parameters of Cs$_2$BI$_6$ (B = Pt, Pd, Te, Sn) vacancy-ordered double perovskites}
\begin{table}[htbp]
	\caption{Calculated lattice parameters (\AA) of Cs$_2$BI$_6$ (B = Pt, Pd, Te, Sn) vacancy-ordered double perovskites using different exchange-correlation ($\varepsilon_{xc}$) functionals.} 
	\begin{center}
		\begin{adjustbox}{width=0.8\textwidth}
			\begin{tabular}[c]{|c|c|c|c|c|} \hline
				$\textbf{Configurations}$ &  $\textbf{Experimental}$ & $\textbf{PBE}$  & $\textbf{PBE-vdW}$ & $\textbf{optB86-vdW}$\\ \hline
				Cs$_2$PtI$_6$  & 11.37\cite{faizan2021electronic}   & 11.74    &   11.47 & 11.29\\ \hline
				Cs$_2$PdI$_6$ & 11.33\cite{schupp2000crystal}     & 11.67    &   11.42 & 11.23\\ \hline
				Cs$_2$TeI$_6$ & 11.70\cite{maughan2016defect}   & 12.06   &   11.87  & 11.65\\ \hline
				Cs$_2$SnI$_6$ & 11.65\cite{maughan2016defect}   & 12.00   &   11.82  & 11.57\\ \hline			
			\end{tabular}
		\end{adjustbox}
		\label{Table1}
	\end{center}
\end{table}
Table \ref{Table1} shows the lattice parameters of vacancy-ordered double perovskites calculated using PBE and optB86 exchange-correlation ($\varepsilon_{xc}$) functionals along with van der Waals (vdW) forces. The two-body vdW interaction, devised by Tkatchenko-Scheffler, has been used during optimization. The optB86-vdW $\varepsilon_{xc}$ functional reproduces the lattice parameters of vacancy-ordered double perovskites close to experimental ones.
\newpage

\section{Goldschmidt tolerance factor ($t$), octahedral factor ($\mu$) and new tolerance factor ($\tau$) of Cs$_2$BI$_6$ (B = Pt, Pd, Te, Sn) vacancy-ordered double perovskites}
Goldschmidt tolerance factor ($t$) and octahedral factor ($\mu$) are given as:
\begin{equation}
	t = \frac{r_A+r_X}{\sqrt{2}(r_B+r_X)}, \: \: \mu=\frac{r_B}{r_X}
\end{equation}
where $r_A$, $r_B$, and $r_X$ are the Shannon ionic radii\cite{shannon1976revised} for $A$, $B$ and $X$ ions, respectively. The Shannon radii for Cs$^+$, Pt$^{4+}$, Pd$^{4+}$, Te$^{4+}$, Sn$^{4+}$ and I$^{-}$ are 1.88, 0.63, 0.62, 0.69 and 0.97\:\AA, respectively. For stable cubic perovskites, the ranges of $t$ and $\mu$ are 0.8 $\leq$ $t$ $\leq$ 1.0 and 0.29 $\leq$ $\mu$ $\leq$ 0.55.
The calculated values given in Table \ref{Table2} show that the considered perovskites are stable in cubic structures.
Recently, Bartel \textit{et al.} have reported a new tolerance factor ($\tau$) to predict the stability of a perovskite, which is given as:
\begin{equation}
	\tau = \frac{r_X}{r_B} - n_A \Biggl( n_A - \frac{r_A/r_B}{ln(r_A/r_B)}\Biggr) 
\end{equation}
where $n$$_A$ is the oxidation state of cation A, $r$$_i$ is the ionic radius of ion $i$ and $r$$_A$ $>$ $r$$_B$ by definition. $\tau$ $<$ 4.18 indicates the formation of perovskite (92\% accuracy). Since the range of $\tau$ is calculated for ABX$_3$ and A$_2$BB$^{'}$X$_6$ double perovskites, this may deviate for vacancy-ordered double perovskites (due to defects). 
\begin{table}[ht!]
	\caption{Goldschmidt tolerance factor ($t$), octahedral factor ($\mu$) and new tolerance factor ($\tau$) of Cs$_2$BI$_6$ vacancy-ordered double perovskites} 
	\begin{center}
		\begin{adjustbox}{width=0.5\textwidth}
			\begin{tabular}[c]{|c|c|c|c|} \hline
				$\textbf{Configurations}$ &  $\textbf{\textit{t}}$ & $\mathbf{\mu}$  & $\mathbf{\tau}$\\ \hline
				Cs$_2$PtI$_6$  & 1.02  &  0.28  &   5.25 \\ \hline
				Cs$_2$PdI$_6$ & 1.02   &  0.28 &   5.31  \\ \hline
				Cs$_2$TeI$_6$ & 0.91   &  0.44 &   4.19 \\ \hline
				Cs$_2$SnI$_6$ & 0.99   & 0.31  &   4.90 \\ \hline			
			\end{tabular}
		\end{adjustbox}
		\label{Table2}
	\end{center}
\end{table}
\clearpage
\newpage

\section{Gibbs free energy ($G$) of Cs$_2$BI$_6$ (B = Pt, Pd, Te, Sn) vacancy-ordered double perovskites}
Reactions for decomposition of Cs$_2$BI$_6$ vacancy-ordered double perovskites:\\
(i) \textrm{Cs$_2$PtI$_6$} $\longrightarrow$ \textrm{2CsI} + \textrm{PtI$_4$} \\
$\Delta H_\textrm{D} = E\mathrm{(Cs_2PtI_6)} - 2E\mathrm{(CsI)} - E\mathrm{(PtI_4)}$\\\
(ii) \textrm{Cs$_2$PdI$_6$} $\longrightarrow$ \textrm{CsI$_3$} + \textrm{PdI$_2$} + \textrm{CsI} \\
$\Delta H_\textrm{D} = E\mathrm{(Cs_2PdI_6)} - E\mathrm{(CsI_3)} - E\mathrm{(PdI_2)} - E\mathrm{(CsI)}$\\
(iii) \textrm{Cs$_2$TeI$_6$} $\longrightarrow$ \textrm{2CsI} + \textrm{TeI$_4$} \\
$\Delta H_\textrm{D} = E\mathrm{(Cs_2TeI_6)} - 2E\mathrm{(CsI)} - E\mathrm{(TeI_4)}$\\
(iv) \textrm{Cs$_2$SnI$_6$} $\longrightarrow$ \textrm{2CsI} + \textrm{SnI$_4$} \\
$\Delta H_\textrm{D} = E\mathrm{(Cs_2SnI_6)} - 2E\mathrm{(CsI)} - E\mathrm{(SnI_4)}$\\

\noindent Here $E$(Cs$_2$BI$_6$), $E$(CsI$_3$), $E$(CsI) and $E$(BI$_4$) are respectively the total DFT energies of Cs$_2$BI$_6$, CsI$_3$, CsI and BI$_4$ and $\Delta H_\textrm{D}$ is the decomposition energy. 
To determine the thermodynamic stability of Cs$_2$BI$_6$ perovskites at room temperature, we have calculated the Gibbs free energy $G$ as a function of temperature $T$ and pressure $P$ given by the following equation: 
\begin{equation}
	G(T,P) = F(T,V) + PV
\end{equation} 

\noindent Here $F(T,V)$ is the Helmholtz free energy as a function of temperature $T$ and volume $V$. To calculate the $PV$ term, we varied the volume of the system and calculated the total energies at different volume. After that, we fitted the data into the empirical equation of state (EOS) to obtain the $E(V)$ function and hence calculated the pressure by $P$ = $-$($\partial E$/$\partial V$)$_T$. According to adiabatic approximation, $F(T,V)$ can be written in terms of ionic vibrational and electronic contributions as follows:
\begin{equation}
	F(T,V) = F_{vib}(T,V)+F_{el}(T,V) \simeq F_{vib}(T,V)+E(T=0\mathrm K,V)
\end{equation}
In the electronic Helmholtz free energy equation, the $TS_{el}$ term is ignored as the electronic temperature effect is negligible for non-metallic systems near room temperature and thus $F_{el}(T,V)=E(T=0\mathrm K,V)-TS_{el} \simeq$ $E(T=0\mathrm K,V)$. Within quasiharmonic approximation (QHA), the ionic Helmholtz free energy $F_{vib}$ can be given as follows:

\begin{equation}
	F_{\textrm{vib}} = 3Mk_\mathrm BT \bigintss_{0}^{\omega_L}ln\:\Biggl\{2\sinh\left[\frac{\hbar\omega(V_0)}{2k_\mathrm BT}\right] \Biggr\} g(\omega)d\omega
\end{equation}
where $M$ is the atomic mass, $\hbar$ is the reduced Planck's constant, $k_\mathrm B$ is the Boltzmann constant and $\omega_i$ is the phonon frequency and $g(w)$ is the normalized phonon DOS. Subsequently, the thermodynamic stability of Cs$_2$BI$_6$ perovskites is estimated by calculating the Gibbs free energy difference as follows: \\
\begin{equation}
	\Delta G (T,P)=G_{\mathrm{Cs_2BI_6}}(T,P)-\big[2\:G_{\mathrm{CsI}}(T,P)+G_{\mathrm{BI_4}}(T,P)\big]
\end{equation}

\begin{figure}[h]
	\includegraphics[width=0.65\textwidth]{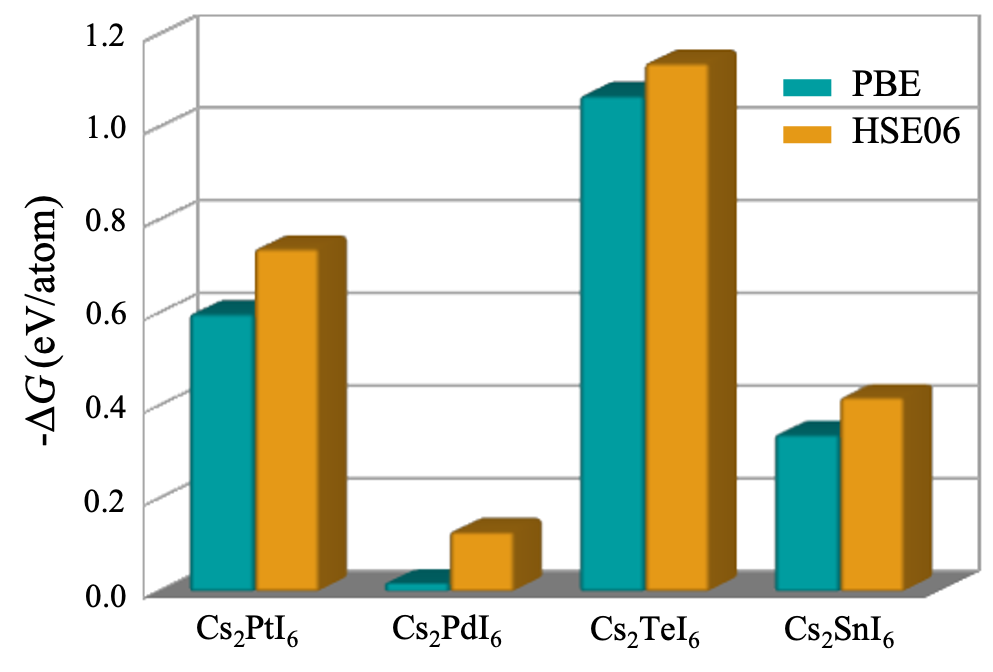}
	\caption{$\Delta G\:(T,P)$ of Cs$_2$BI$_6$  vacancy-ordered double perovskites calculated using PBE and HSE06 $\varepsilon_{xc}$ functionals.}
	\label{d}
\end{figure}

\noindent The $\Delta G(T,P)$ values are listed in Table \ref{Table3}. The negative values confirm the thermodynamic stability of these vacancy-ordered double perovskites (see Figure \ref{d}).
\begin{table}[htbp]
	\caption{$\Delta G(T,P)$ of Cs$_2$BI$_6$ perovskites calculated using PBE and HSE06 $\varepsilon_{xc}$ functionals} 
	\begin{center}
		\begin{adjustbox}{width=0.85\textwidth}
			\begin{tabular}[c]{|c|c|c|c|} \hline
				$\textbf{Configurations}$ &  $\mathbf {\Delta G}$ $\mathbf{(eV/atom) (PBE)}$ &  $\mathbf {\Delta G}$ $\mathbf{(eV/atom) (HSE06)}$  \\ \hline
				Cs$_2$PtI$_6$  & -0.62  & -0.77   \\ \hline
				Cs$_2$PdI$_6$ & -0.04  & -0.16    \\ \hline
				Cs$_2$TeI$_6$ & -1.09   &  -1.17      \\ \hline
				Cs$_2$SnI$_6$ & -0.35  & -0.44   \\ \hline			
			\end{tabular}
		\end{adjustbox}
		\label{Table3}
	\end{center}
\end{table}
\clearpage
\newpage

\section{Mechanical properties of Cs$_2$BI$_6$ (B = Pt, Pd, Te, Sn) vacancy-ordered double perovskites}
The Voigt bulk (\textit{B$_V$}) and shear (\textit{G$_V$}) moduli, Reuss bulk (\textit{B$_R$}) and shear (\textit{G$_R$}) moduli  are calculated using the following relations:
\begin{equation}
	B_V = B_R = \frac{(C_{11} + 2 C_{12})}{3}
\end{equation}
\begin{equation}
	G_V = \frac{(C_{11} - C_{12} + 3 C_{44})}{5} 
\end{equation}
\begin{equation}
	G_R = \frac{5(C_{11} - C_{12}) C_{44}}{4C_{44} + 3 (C_{11} - C_{12})}
\end{equation}

According to Voigt-Reuss-Hill approximations\cite{voigt2016lehrbuch}, Young's modulus ($E$) and Poissons's ratio ($\nu$) are obtained as:
\begin{equation}
 B= \frac{B_V+B_R} {2} , \: \: \: \: \: \:  G= \frac{G_V+G_R} {2}  
\end{equation}
\begin{equation}
	E =  \frac{9BG}{3B+G}
\end{equation}

\begin{equation}
\nu = \frac{3B-2G}{3B+G}
\end{equation}

\noindent The ductility of these perovskites is studied in terms of Pugh's ($B/G$) and Poisson's ratio ($\nu$). If the $B/$G is found to be greater (or lower) than 1.75, the material is ductile (or brittle). For $\nu$, the limiting value is 0.26. The calculated values of $B/G$ and  $\nu$ show that the studied vacancy-ordered double perovskites are ductile. 
\newpage

\section{Band structures of Cs$_2$BI$_6$ (B = Pt, Pd, Te, Sn) vacancy-ordered double perovskites}
\begin{figure}[h]
	\includegraphics[width=1.0\textwidth]{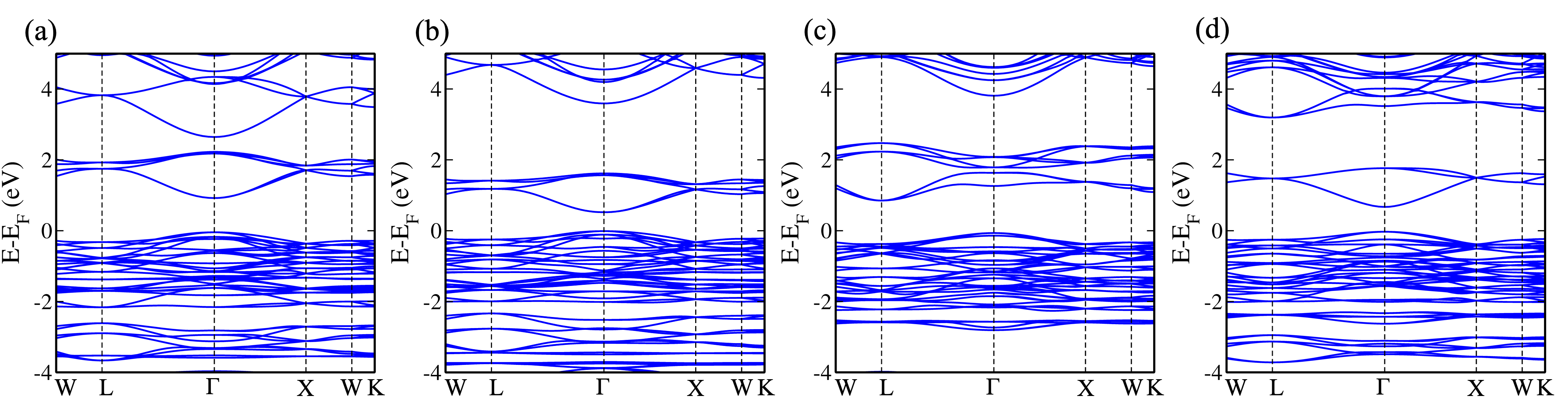}
	\caption{Band structures of (a) Cs$_2$PtI$_6$, (b) Cs$_2$PdI$_6$, (c) Cs$_2$TeI$_6$ and (d) Cs$_2$SnI$_6$  vacancy-ordered double perovskites, calculated using HSE06+SOC $\varepsilon_{xc}$ functional.}
	\label{d1}
\end{figure}
The band structures for Cs$_2$PtI$_6$, Cs$_2$PdI$_6$, Cs$_2$TeI$_6$ and Cs$_2$SnI$_6$ are calculated using HSE06+SOC $\varepsilon_{xc}$ functional (see Figure \ref{d1}) and the band gaps of these perovskites are 0.96, 0.51, 1.49 and 0.70 eV, respectively.  

\begin{figure}[htp]
\includegraphics[width=0.65\textwidth]{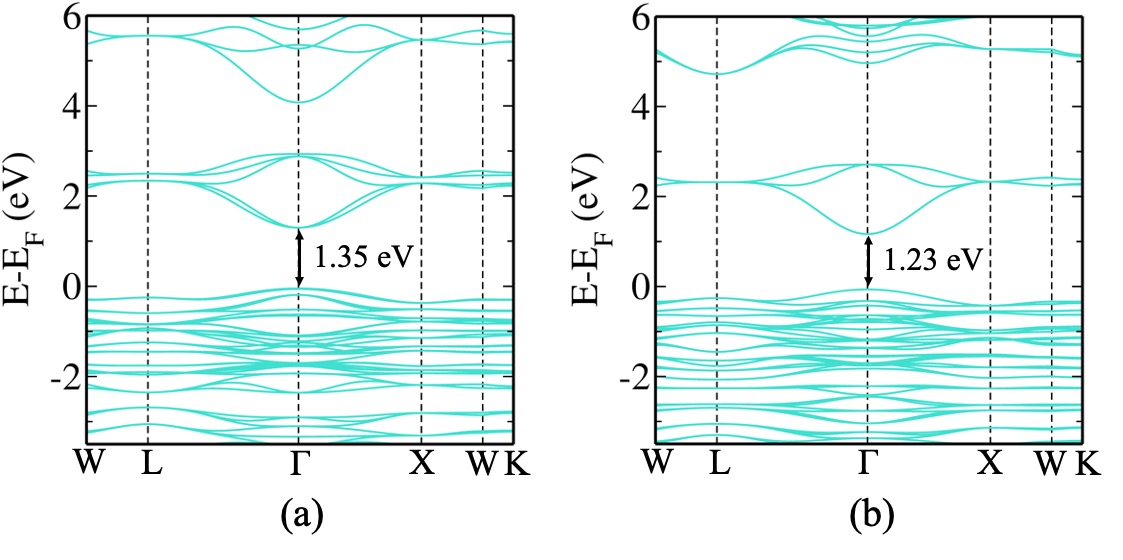}
\caption{Band structures of (a) Cs$_2$PtI$_6$ and (b) Cs$_2$SnI$_6$ vacancy-ordered double perovskites, calculated using G$_0$W$_0$@PBE+SOC.}
\label{l}
\end{figure}
The band structures for Cs$_2$PtI$_6$ and Cs$_2$SnI$_6$ are calculated using G$_0$W$_0$@PBE+SOC $\varepsilon_{xc}$ functional (see Figure \ref{l}) and the band gaps of these perovskites are 1.35 and 1.23 eV, respectively.  
\newpage

\section{Projected density of states (pDOS) of Cs$_2$BI$_6$ (B = Pt, Pd, Te, Sn) vacancy-ordered double perovskites}
\begin{figure}[h]
	\includegraphics[width=1.0\textwidth]{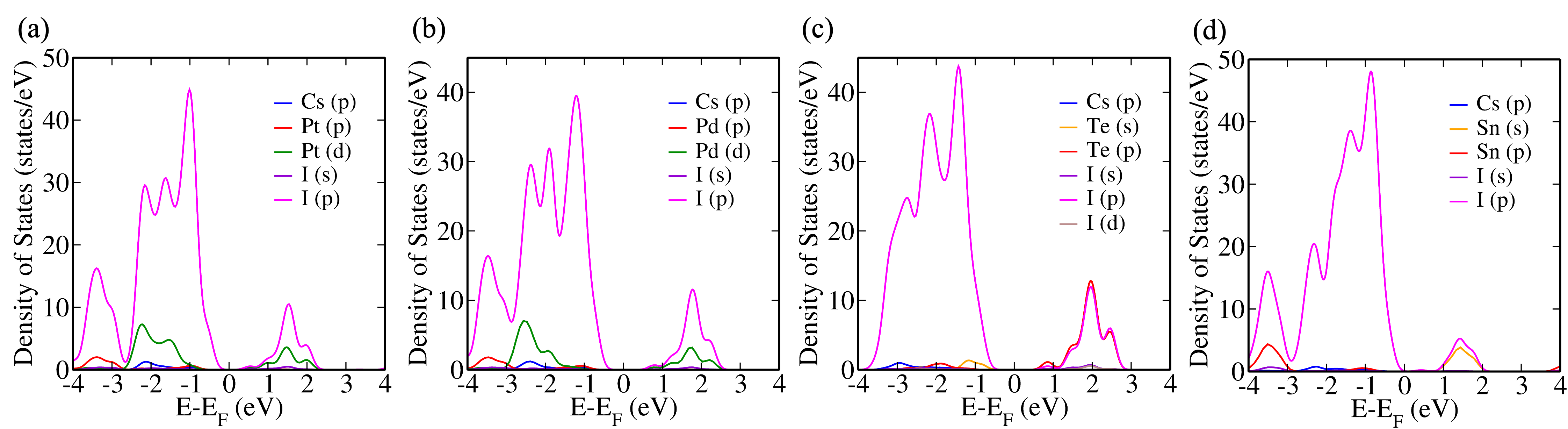}
	\caption{Projected density of states (pDOS) of (a) Cs$_2$PtI$_6$, (b) Cs$_2$PdI$_6$, (c) Cs$_2$TeI$_6$ and (d) Cs$_2$SnI$_6$  vacancy-ordered double perovskites, calculated using HSE06+SOC $\varepsilon_{xc}$ functional.}
	\label{d2}
\end{figure}
Figure \ref{d2} shows the pDOS of all four vacancy-ordered double perovskites. The valence band maximum (VBM) of these perovskites are mostly dominated by I-$p$ orbitals, while conduction band minimum (CBm) are contributed by I-$p$ orbitals along with Pt-$d$, Pd-$d$, Te-$p$ and Sn-$s$ orbitals in Cs$_2$PtI$_6$, Cs$_2$PdI$_6$, Cs$_2$TeI$_6$ and Cs$_2$SnI$_6$, respectively. The VBM is also composed of Pt-$d$, Pd-$d$ and Te-$s$ orbitals in Cs$_2$PtI$_6$, Cs$_2$PdI$_6$ and Cs$_2$TeI$_6$, respectively. 
There is a strong hybridization of Te-$p$ and I-$p$ orbitals in Cs$_2$TeI$_6$ and Sn-$s$ and I-$p$ orbitals in Cs$_2$SnI$_6$.
\newpage

\section{$k$-grid convergence in Cs$_2$BI$_6$ (B = Pt, Pd, Te, Sn) vacancy-ordered double perovskites}
\noindent For the reliability of the calculations, we have checked the convergence of the properties viz., energy ($E$), Seebeck coefficient ($S$), imaginary (Im($\varepsilon$)) and real (Re($\varepsilon$)) part of the dielectric function in the vacancy-ordered double perovskites with \textit{k}-grid. The energies of the Cs$_2$BI$_6$ perovskites in Table \ref{E} show that 4$\times$4$\times$4 $k$-grid is sufficient to calculate $E$. Also, the values of $S$ with 4$\times$4$\times$4 and 6$\times$6$\times$6 $k$-grids are nearly similar and are listed in Table \ref{S}.
\begin{table}[htbp]
	\caption {Energies (eV) of Cs$_2$BI$_6$ (B = Pt, Pd, Te, Sn) vacancy-ordered double perovskites at different \textit{k}-grids calculated using PBE $\varepsilon_{xc}$ functional.}
	\begin{center}
		\begin{adjustbox}{width=0.75\textwidth} 
			\setlength \extrarowheight {+5pt}
			\begin{tabular}[c]{|c|c|c|c|c|c|} \hline		
				& \multicolumn{5} {c|} {\textbf{\textit{E} (eV)}} \\ \hline
				\textbf{Configurations} & \textbf{2$\times$2$\times$2} & \textbf{3$\times$3$\times$3} & \textbf{4$\times$4$\times$4} & \textbf{5$\times$5$\times$5} & \textbf{6$\times$6$\times$6}  \\ \hline
				Cs$_2$PtI$_6$        &  -110.5206  & -110.5481 & -110.5493 & -110.5493  & -110.5494  \\ \hline
				Cs$_2$PdI$_6$       & -106.3718   &  -106.4057  &  -106.4056  & -106.4057  & -106.4052  \\ \hline
				Cs$_2$TeI$_6$       &  -98.7717  & -98.7872 & -98.7873 & -98.7874   &   -98.7873 \\ \hline
				Cs$_2$SnI$_6$       &  -105.2440  &  -105.2542 & -105.2547 & -105.2550  &  -105.2532 \\ \hline
			\end{tabular}
		\end{adjustbox}
		\label{E}
	\end{center}
\end{table}

\begin{table}[htbp]
	\caption {Seebeck coefficient (\textmu V/K) of Cs$_2$BI$_6$ (B = Pt, Pd, Te, Sn) vacancy-ordered double perovskites at different \textit{k}-grids calculated using HSE06 $\varepsilon_{xc}$ functional.}
	\begin{center}
		\begin{adjustbox}{width=0.38\textwidth} 
			\setlength \extrarowheight {+5pt}
			\begin{tabular}[c]{|c|c|c|} \hline		
				& \multicolumn{2} {c|} {\textbf{\textit{S} (\textmu V/K)}} \\ \hline
				\textbf{Configurations} & \textbf{4$\times$4$\times$4} & \textbf{6$\times$6$\times$6}  \\ \hline
				Cs$_2$PtI$_6$        &  710.01  & 710.06  \\ \hline
				Cs$_2$PdI$_6$       & 148.23  &  148.31  \\ \hline
				Cs$_2$TeI$_6$       &  190.11  & 190.18 \\ \hline
				Cs$_2$SnI$_6$       &  290.52  &  290.59 \\ \hline
			\end{tabular}
		\end{adjustbox}
		\label{S}
	\end{center}
\end{table}

\begin{figure}[h]
	\includegraphics[width=0.8\textwidth]{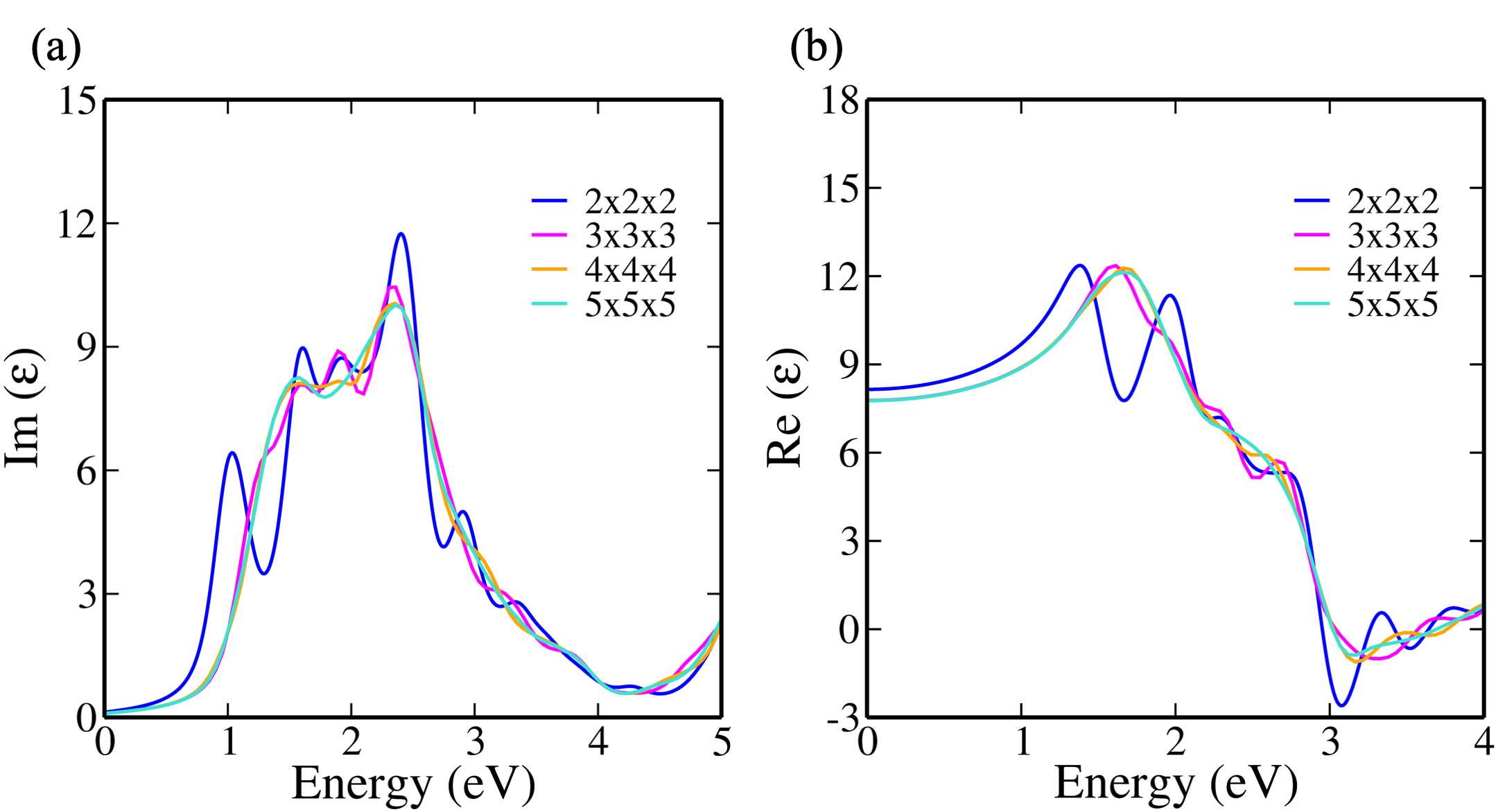}
	\caption{Imaginary (Im($\varepsilon$)) and real (Re($\varepsilon$)) part of the dielectric function for Cs$_2$TeI$_6$ vacancy-ordered double perovskite calculated using PBE $\varepsilon_{xc}$ functional.}
	\label{k}
\end{figure}

\noindent We have also checked the \textit{k}-grid convergence for imaginary (Im($\varepsilon$)) and real part (Re($\varepsilon$)) of the dielectric function. Figure \ref{k} (a) and (b) show the variation of Im($\varepsilon$) and Re($\varepsilon$) for Cs$_2$TeI$_6$ vacancy-ordered double perovskite. On increasing the \textit{k}-grid, no shift is observed in the first peak position of Im($\varepsilon$) part of dielectric function. Hence, the 4$\times$4$\times$4 \textit{k}-grid is sufficient to compute quasiparticle (G$_0$W$_0$) band gap. Therefore, we have calculated all the properties with 4$\times$4$\times$4 $k$-grid except when stated otherwise.\\
 
\newpage

\section{Computational details for calculation of Seebeck coefficient ($S$) and thermoelectric figure of merit ($zT$) in Cs$_2$BI$_6$ (B = Pt, Pd, Te, Sn) vacancy-ordered double perovskites}
 The thermoelectric properties viz., Seebeck coefficient ($S$) and thermoelectric figure of merit ($zT$) are calculated using the Boltzmann transport equation and the rigid band approach as implemented in the BoltzTraP code\cite{madsen2006boltztrap}. The cutoff energy of 600 eV is used for the plane wave basis set such that the total energy calculations are converged within 10$^{-5}$ eV. The $\Gamma$-centered 4$\times$4$\times$4 $k$-grid is used to sample the Brillouin zone after checking the convergence with $k$-grids (see Table \ref{P}). Seebeck coefficient ($S$) is determined as a function of chemical potential ($\mu$) for Cs$_2$BI$_6$ perovskites. After that, we have plotted $zT$ as a function of temperature ($T$). The ``average $zT$" is calculated by taking the mean of all $zT$ values at different temperatures. 
 \begin{table}[htbp]
 	\caption {Seebeck coefficient ($S$) of Cs$_2$BI$_6$ (B = Pt, Pd, Te, Sn) vacancy-ordered double perovskites at different \textit{k}-grids calculated using HSE06 $\varepsilon_{xc}$ functional.}
 	\begin{center}
 		\begin{adjustbox}{width=0.38\textwidth} 
 			\setlength \extrarowheight {+5pt}
 			\begin{tabular}[c]{|c|c|c|} \hline		
 				& \multicolumn{2} {c|} {\textbf{\textit{S} (\textmu V/K)}} \\ \hline
 				\textbf{Configurations} & \textbf{4$\times$4$\times$4} & \textbf{6$\times$6$\times$6}  \\ \hline
 				Cs$_2$PtI$_6$        &  710.01  & 710.06  \\ \hline
 				Cs$_2$PdI$_6$       & 148.23  &  148.31  \\ \hline
 				Cs$_2$TeI$_6$       &  190.11  & 190.18 \\ \hline
 				Cs$_2$SnI$_6$       &  290.52  &  290.59 \\ \hline
 			\end{tabular}
 		\end{adjustbox}
 		\label{P}
 	\end{center}
 \end{table}

\newpage
\section{Thermal conductivity ($\kappa$) of Cs$_2$BI$_6$ (B = Pt, Pd, Te, Sn) vacancy-ordered double perovskites}
\begin{figure}[h]
	\includegraphics[width=0.99\textwidth]{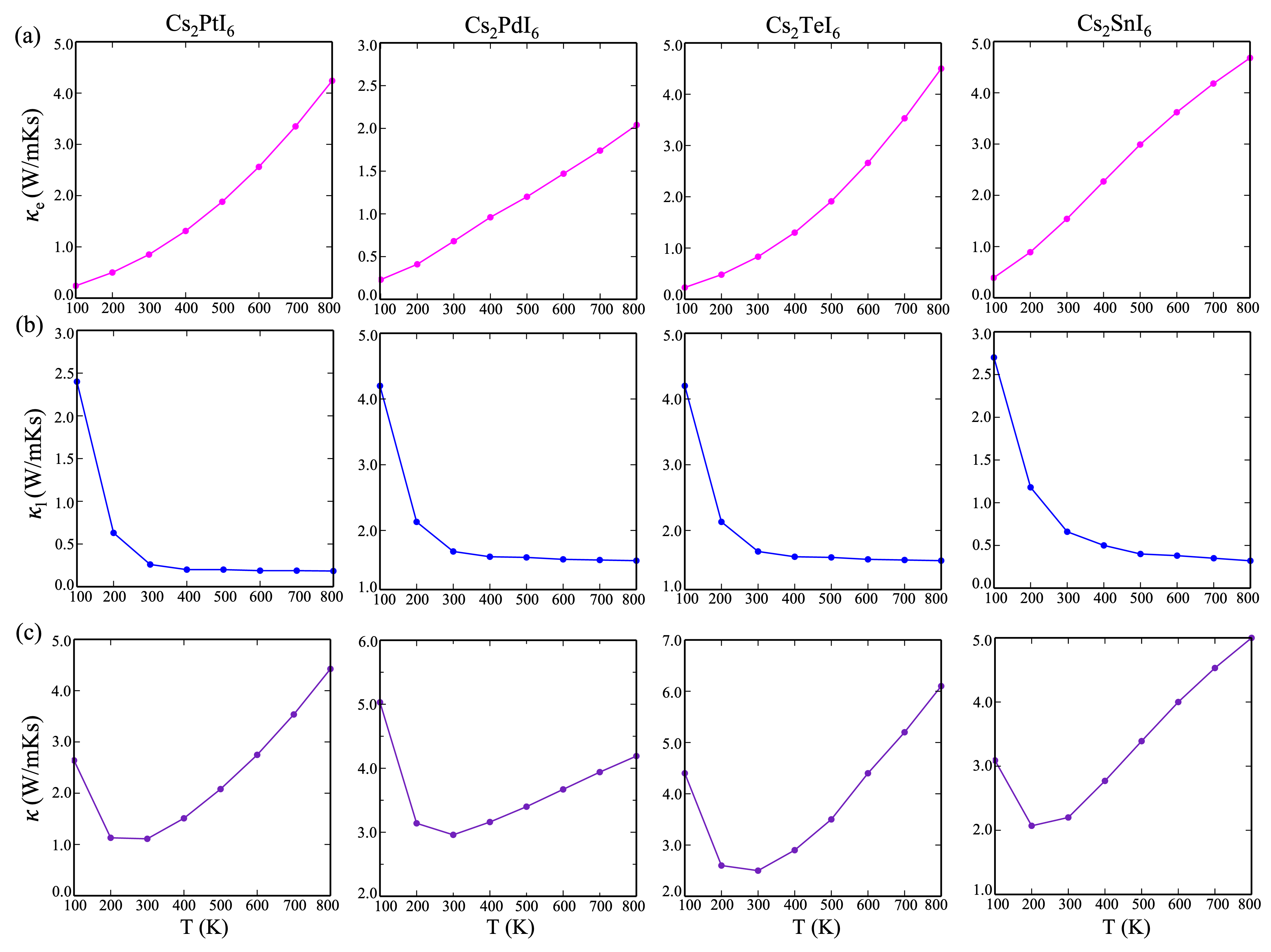}
	\caption{Calculated (a) electronic thermal conductivity, (b) lattice thermal conductivity and (c) total thermal conductivity in Cs$_2$BI$_6$ vacancy-ordered double perovskites, calculated using PBE $\varepsilon_{xc}$ functional.}
	\label{cond}
\end{figure}
\newpage
\section{Calculation of harmonic (U$_{h}$) and anharmonic energies (U$_{ah}$) in Cs$_2$BI$_6$ (B = Pt, Pd, Te, Sn) vacancy-ordered double perovskites}
Harmonic energy (U$_{h}$) is calculated as per the following equation:
\begin{equation}
	U_h= \sum_{i} \frac{h\nu_{i}}{2}+\sum_{i}{k}_{\textrm{B}}T~{\textrm{ln}}\left[1-\exp\left(\frac{h\nu_{i}}{k_{\textrm{B}}T}\right)\right]
\end{equation}
where $k$$_\textrm{B}$, $h$, and $T$ are respectively the Boltzmann constant, Planck's constant, and temperature. $\nu_{i}$ represent frequencies of vibration, which we get from vibration calculation.\\
To calculate the anharmonic energy, we have performed AIMD calculations at 250, 500, 750, 1000, 1250, 1500, 1750 and 2000\:K temperatures using Nose-Hoover thermostat with time and time-step being 8 ps and 1 fs, respectively. This data is then supplied to a post-processing python package pyHMA, which determines the anharmonic energy according to the relation:
\begin{equation}
	U_{ah}= \langle U +\frac{1}{2}\mathbf{F}.\Delta \mathbf{r} \rangle - U_{lat}
\end{equation}
where the \textbf{F} vector represents the forces on all atoms, and $\Delta$\textbf{r} is the displacement of each atom from its lattice (equilibrium) site. The pyHMA package reads MD data from output file (vasprun.xml) to compute anharmonic energy. The package can be downloaded from the development version on GitHub.
\newpage

\section{Effective mass calculation of Cs$_2$PtI$_6$ vacancy-ordered double perovskite}
Calculation of effective masses using SUMO package.\\
\textbf{Electron effective masses:} \\
$m$$_e$ : 0.491 $|$ band 284 $|$ [0.00, 0.00, 0.00] ($\Gamma$) → [0.50, 0.50, 0.50] (L)\\ 
$m$$_e$ : 0.489 $|$ band 284 $|$ [0.00, 0.00, 0.00] ($\Gamma$) → [0.50, 0.00, 0.50] (X)\\
$m$$_e$ : 0.491 $|$ band 285 $|$ [0.00, 0.00, 0.00] ($\Gamma$) → [0.50, 0.50, 050] (L)\\
$m$$_e$ : 0.489 $|$ band 285 $|$ [0.00, 0.00, 0.00] ($\Gamma$) → [0.50, 0.00, 0.50] (X)\\
$m$$_e$ : 0.491 $|$ band 286 $|$ [0.00, 0.00, 0.00] ($\Gamma$) → [0.50, 0.50, 0.50] (L)\\
$m$$_e$ : 0.489 $|$ band 286 $|$ [0.00, 0.00, 0.00] ($\Gamma$) → [0.50, 0.00, 0.50] (X)\\
To calculate the average electron mass at $\Gamma$-point, we have taken the average of masses along $\Gamma$ → L and $\Gamma$ → X directions.\\
Average effective electron mass = 0.49 m$_0$
\newpage

\section{Calculation of longitudinal optical phonon frequency ($\omega_{\textrm{LO}}$), effective mass of polaron (\textit{m}$_\textrm{P}$), polaron radii (\textit{l}$_\textrm{P}$) and polaron mobility ($\mu_\textrm{P}$)}
The average frequency of the LO (Longitudinal Optical) modes is calculated from infrared (IR) phonon frequencies obtained using the density functional perturbation theory (DFPT). The free electrons interact with lattice, which consists of more than one infrared-active optical phonon mode. To get an effective longitudinal optical phonon frequency ($\omega$$_{\textrm{LO}}$), we have used the ansatz of Hellwarth and Biaggio\cite{hellwarth1999mobility}. The oscillator strength \textit{W}$_i$ of the LO phonon modes is calculated using the LO-TO (Transverse Optical) splitting as given by the following equation:\\
\begin{equation}
	\begin{split}
		W_i^2=\frac{1}{\varepsilon_\infty}(\omega^2_{\textrm{LO},i}-\omega^2_{\textrm{TO},i})
		\label{eq8}
	\end{split}
\end{equation}
with subsequent quadratic mean (W$^2$ = $\sum$W$_i^2$) over single oscillator strengths to obtain the oscillator strength of the single phonon branch W. The frequency $\omega$$_{\textrm{LO}}$ (in wave numbers) of this branch is then calculated by solving
\begin{equation}
	\begin{split}
		\frac{W^2}{\omega_{\textrm{LO}}}\textrm{cot}\:h\left(\frac{hc\:\omega_{\textrm{LO}}}{2k_BT}\right)=\sum_{i=1}^{2}\frac{W^2_i}{\omega_{\textrm{LO},i}}\textrm{cot}\:h\left(\frac{hc\:\omega_{\textrm{LO},i}}{2k_BT}\right)
		\label{eq9}
	\end{split}
\end{equation}
Further, within the Fr\"{o}hlich's polaron theory, as extended by Feynman, the effective mass of polaron ($m$$_\textrm{p}$)~\cite{PhysRev.97.660} can be calculated as:
\begin{equation}
	\begin{split}
		m_{\textrm{P}} = m^*\left(1+\frac{\alpha}{6} +  \frac{\alpha^2}{40} + ......\right)
		\label{eq4}
	\end{split}
\end{equation}
where $m$$^{*}$ is the effective mass of electron in terms of rest mass of electron $m_0$. The polaron radius ($l$$_\textrm{P}$)\cite{sendner2016optical} can be determined as:
\begin{equation}
	\begin{split}
		l_\textrm{P} = \sqrt{\frac{h}{2cm^*\omega_{\textrm{LO}}}}
		\label{eq5}
	\end{split}
\end{equation} 
where $h$ is the Planck's constant and $c$ is the speed of light.
Now, according to Hellwarth polaron model\cite{hellwarth1999mobility}, the polaron mobility ($\mu_{\textrm{P}}$) is defined as follows:
\begin{equation}
	\begin{split}
		\mu_{\textrm{P}} = \frac{(3\sqrt{\pi}e)}{2\pi {c}\:\omega_{\textrm{LO}}m^*\alpha} \frac{\textrm{sin}\:h{(\beta/2)}}{\beta^{5/2}}\frac{w^3}{v^3}\frac{1}{\textit{K}}
		\label{eq5}
	\end{split}
\end{equation}
where, $\beta$ = $hc$\:$\omega_{\textrm{LO}}$/$k_{\textrm{B}}$T, \textit{e} is the electronic charge, $m$$^*$ is the effective mass of charge carrier, $\textit{w}$ and $\textit{v}$ correspond to temperature dependent variational parameters. $\textit{K}$ is a function of $\textit{v}$, $\textit{w}$, and $\beta$ defined as follows: 
\begin{equation}
	\begin{split}
		K(a,b) = \int_0^\infty du\:[u^2 + a^2 - b\:\textrm{cos}(vu)]^{-3/2}\textrm{cos}(u)
		\label{eq12}
	\end{split}
\end{equation}
Here, $a$$^2$ and $b$ are calculated as: 
\begin{equation}
	\begin{split}
		a^2 = (\beta /2)^2 + \frac{(v^2 - w^2)}{w^2v} \beta\:\textrm{cot}\:h(\beta v/2)
		\label{eq13}
	\end{split}
\end{equation} 
\begin{equation}
	\begin{split}
		b =  \frac{(v^2 - w^2)}{w^2v}\frac{\beta}{\textrm{sin}\:h(\beta v/2)}
		\label{eq14}
	\end{split}
\end{equation}  
\newpage

\providecommand{\latin}[1]{#1}
\makeatletter
\providecommand{\doi}
{\begingroup\let\do\@makeother\dospecials
	\catcode`\{=1 \catcode`\}=2 \doi@aux}
\providecommand{\doi@aux}[1]{\endgroup\texttt{#1}}
\makeatother
\providecommand*\mcitethebibliography{\thebibliography}
\csname @ifundefined\endcsname{endmcitethebibliography}
{\let\endmcitethebibliography\endthebibliography}{}
